\documentclass{aip-cp}

\usepackage[numbers]{natbib}
\usepackage{placeins}
\usepackage{rotating}
\usepackage{lineno}

\begin{document}

\title{  Ultra-peripheral Collisions with the ATLAS Detector}

\author{Miguel Arratia$^{1}$ for the ATLAS collaboration}
\affil[aff1]{Cavendish Laboratory, University of Cambridge.}

\maketitle

\begin{abstract}
The first ATLAS measurement of ultra-peripheral collisions is described. Dimuon pairs from photon--photon scattering have been measured with data of Pb+Pb collisions at \mbox{$\sqrt{s_{\mathrm{NN}}} = 5.02$ TeV} taken with the ATLAS detector at the LHC. The measured cross-section is presented as a function of dimuon invariant mass and rapidity and is well described by STARLIGHT 1.1 calculations. This measurement paves the way for new studies of photon-induced reactions at the LHC. 
\end{abstract}

\section{INTRODUCTION}
Collisions of nuclear beams with relativistic energies are normally used to produce hot and dense quark-gluon matter. But the large electromagnetic fields ($\approx 10^{14}$ T) generated by the nuclear beams also provide access to photon-induced reactions. These can be studied in ``ultra-peripheral collisions" in which the impact parameter is large enough to suppress strong interactions. The maximum energy of the quasi-real photons coherently emitted by Pb nuclei at the LHC is about \mbox{75 GeV}, which is 25 times larger than at RHIC.\footnote{The photon energy is exponentially suppressed beyond $\gamma/R$. Here $\gamma$ is the Lorentz factor of the Pb beam and $R$ is the nuclear radius.} The measurement presented here is of the differential cross-section of muon pairs produced in photon--photon scattering~\cite{ATLAS:2016vdy}.   
\section{ATLAS DETECTOR}
The ATLAS detector at the LHC covers nearly 4$\pi$ around the collision point.\footnote{ATLAS uses a right-handed coordinate system with its origin at the nominal interaction point in the center of the detector, the $z$-axis points along the beam pipe, and the $y$-axis points upward. Cylindrical coordinates $(r,\phi)$ are used in the transverse plane, $\phi$ being the azimuthal angle around the beam pipe. The pseudorapidity is defined in terms of the polar angle $\theta$ as $\eta=-\ln\tan(\theta/2)$.} It consists of an inner tracking detector surrounded by a thin superconducting solenoid, electromagnetic and hadronics calorimeters, forward scintillators, and a muon spectrometer incorporating superconducting toroid magnets. The ATLAS trigger system is implemented in hardware (Level-1) and in software (HLT level). A detailed description of the ATLAS detector is found in~\cite{Aad:2008zzm}. 
\section{DATA AND SIMULATION}
The data used in this measurement were collected during the 2015  \mbox{$\sqrt{s_{\mathrm{NN}}} = 5.02$} TeV Pb+Pb run of the LHC. The data sample corresponds to an integrated luminosity of \mbox{515 $\mu$b$^{-1}$}. The STARLIGHT 1.1 Monte Carlo (MC)  generator~\cite{Starlight} is used to simulate the Pb + Pb $\to \mu^{+}\mu^{-}$ + Pb$^{*}$ + Pb$^{*}$ reaction. This simulation is based on the leading-order QED calculation for $\gamma\gamma\to\mu^{+}\mu^{-}$, and the per-nucleus photon-flux estimated with the ``equivalent photon approximation". The same reconstruction algorithms are applied to data and simulation. 
\section{TRIGGER}
This analysis is based on a single-muon trigger with veto of additional activity in the detector. The Level-1 required at least one track in the muon spectrometer and less than \mbox{50 GeV} of transverse energy in the calorimeters. The HLT level rejected events with more than one hit in the forward scintillators and required an inner detector track with transverse momentum $p_{\mathrm{T}}$ above \mbox{400 MeV}. A total of 248095 events were selected with this trigger.
\section{EVENT SELECTION}
The events were required to have a primary vertex formed solely from two oppositely-charged muons. Both muons were required to satisfy the ``tight" criteria~\cite{Aad:2016jkr}. One of the muons was required to be compatible with the trigger particle, i.e.~to be within an angular distance $\Delta R = \sqrt{\Delta\eta^{2} + \Delta\phi^{2}}<0.5$ of a Level-1 muon-spectrometer track. To ensure reasonable reconstruction efficiency, both muons were required to have \mbox{$p_{\mathrm{T}}>$ 4 GeV} and $|\eta|<2.4$. After all these criteria were applied, 12069 events were selected. 
\section{ANALYSIS OF ACOPLANARITY}
The nuclear form factor limits the $p_{\mathrm{T}}$ of the emitted photons to less than about \mbox{200 MeV}, so the $\gamma\gamma\to\mu^{+}\mu^{-}$ process yields a back-to-back topology in azimuth~\cite{Baltz:2009jk}. The observed $|\Delta \phi|$ distribution shows a peak at $|\Delta \phi|\approx\pi$ with a width consistent with detector resolution, and is well described by the MC simulation. A tail at a few-percent level of the peak extends to lower values of $|\Delta \phi|$ and is not described by the MC simulation. Only events with $|\Delta\phi|>\pi(1-0.008)$ are selected. This cut has an efficiency of 99.9\% for the signal according to the MC simulation. About 5--8\% of the total number of events have $|\Delta\phi|<\pi(1-0.008)$. Some of these events result from strong interactions and thus are a background. However, as shown in more detail in~\cite{ATLAS:2016vdy}, some of these events may also be produced by final-state QED radiation.\footnote{Given that the STARLIGHT calculation includes only leading-order QED calculations, the signal fraction that extends beyond the \mbox{$|\Delta\phi|<\pi(1-0.008)$} cut has not been fully estimated. This is taken into account in the section of systematic uncertainties.} The background contamination is estimated by parametrizing the low-$\Delta\phi$ tail with an exponential function and extrapolating to the signal region. This yields an estimate of 2--4\% depending on the dimuon invariant mass $M_{\mu\mu}$ and rapidity $Y_{\mu\mu}$.
\section{CORRECTIONS}
Event-by-event weights are applied to take into account trigger, muon reconstruction, and vertex efficiency. The trigger and reconstruction efficiency depend strongly on muon $p_{\mathrm{T}}$ and $|\eta|$, as described in~\cite{Aad:2016jkr}, but typically are higher than 80\% and 90\% respectively. The vertex efficiency is 96.8\% according to simulation. The corrected $M_{\mu\mu}$ and $Y_{\mu\mu}$ spectra are divided by the integrated luminosity of the sample to obtain differential cross-sections.
\section{SYSTEMATIC UNCERTAINTIES}
The sources of systematic uncertainty on the differential cross-section measurement are listed below.
\begin{enumerate}
\item {\bf Luminosity}. The uncertainty on the absolute luminosity scale of the 2015 Pb+Pb data is $7\%$.  
\item {\bf Trigger efficiency}. The trigger efficiency uncertainty is estimated by the difference of two independent data-driven methods. This corresponds to a variation of 2\% or less depending on $M_{\mu\mu}$ and $Y_{\mu\mu}$. 
\item {\bf Muon reconstruction}. The muon reconstruction efficiency uncertainty is estimated in reference~\cite{Aad:2016jkr}. It corresponds to about 2--5\% uncertainty in the cross-section measurement depending on  $M_{\mu\mu}$ and $Y_{\mu\mu}$. 
\item {\bf Background estimation}. The nominal background estimation assumes that all the acoplanarity tail is due to background. The other extreme assumption is that it is due to final-state QED radiation. The difference between these methods is taken as a systematic uncertainty. The magnitude of the uncertianty is below $6\%$ for $Y_{\mu\mu}<1.6$ and increases to around 10\% for the $M_{\mu\mu}>$ 15 GeV and $1.6<|Y_{\mu\mu}|<2.4$ interval.
\item {\bf Monte Carlo self-consistency}. The generated dimuon spectrum is compared with the reconstructed spectrum after correction from MC-derived reconstruction and vertex efficiency. The maximum discrepancy is 2\% and is assigned as an overall uncertainty.
\item {\bf Pile-up}. The maximum pile-up rate of the 2015 Pb+Pb run reached 0.5\%. No correction is applied but a systematic uncertainty of 0.5\% is assigned to account for this.
\end{enumerate}

\section{RESULTS}
\begin{figure}[h]
  \centerline{\includegraphics[width=1.0\textwidth]{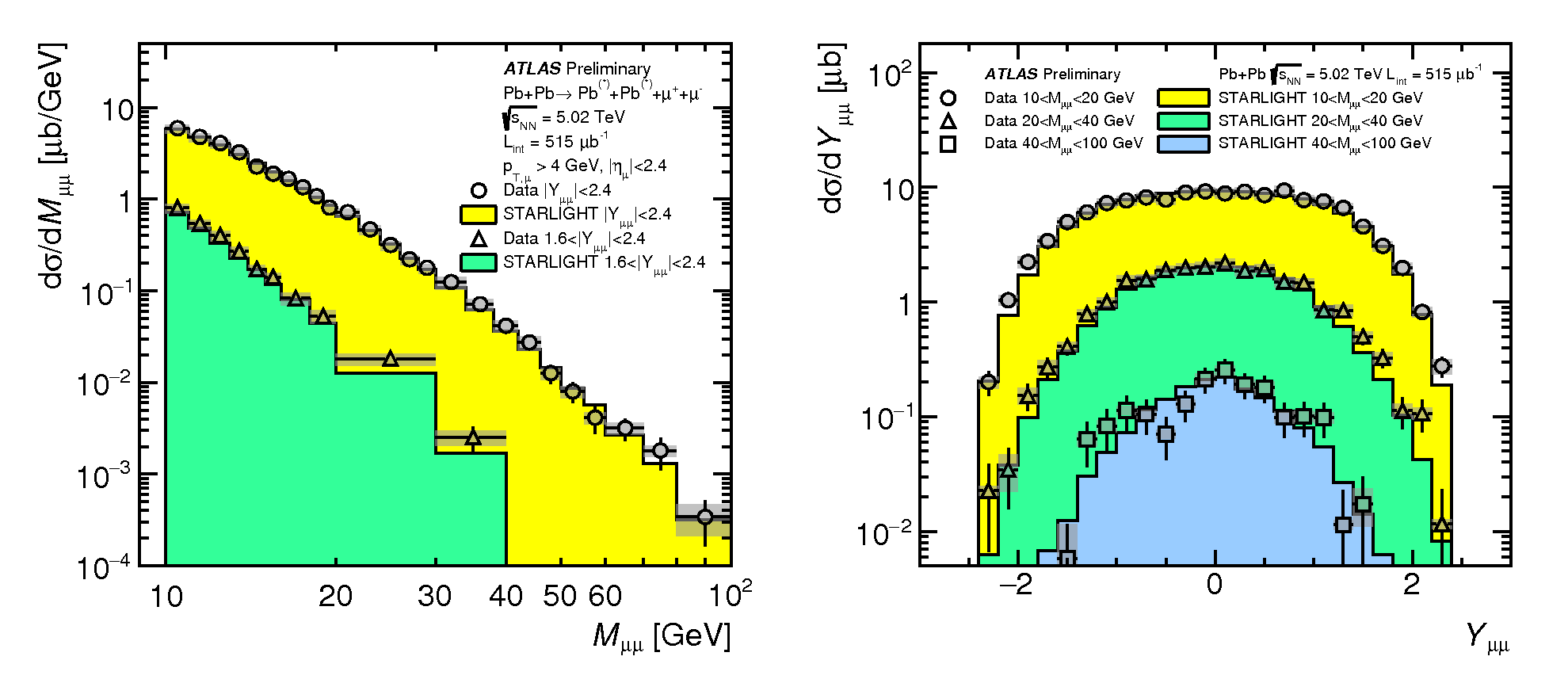}}
  \caption{Differential cross-section of exclusive dimuon production in ultra-peripheral Pb+Pb collisions as a function of pair mass (left) and rapidity (right). The data are shown by symbols while MC predictions are shown as solid histograms. Error bars indicate the statistical uncertainties while the grey bands indicate the total systematic uncertainty. Source: reference~\cite{ATLAS:2016vdy}\label{result}}
\end{figure}

Figure~\ref{result} shows the measured double-differential cross-section and the MC predictions. The STARLIGHT 1.1 simulation describes well the magnitude and shape of all measurements. Integrating the distributions in Figure~\ref{result} gives the total cross-section in the fiducial acceptance. This gives $\sigma(Pb+Pb\to\mu^{+}\mu^{-}+Pb^{*}+Pb^{*}) = 32.2\pm0.3~(\mathrm{stat.})^{+4.0}_{-3.4}~(\mathrm{syst.})~\mu$b. The STARLIGHT 1.1 cross-section is $31.64\pm0.04~(\mathrm{stat.})~\mu$b, well within the stated experimental uncertainties. 
\FloatBarrier
\section{CONCLUSION}
The first preliminary measurement of high-mass exclusive dimuon events in ultra-peripheral heavy-ion collisions is presented. The data corresponds to lead-lead collisions at 5 TeV taken with the ATLAS detector at the LHC. The kinematic coverage of the measurement is larger than any previous measurement of its kind (with nuclear beams), with dimuon rapidity out to $|Y_{\mu\mu}|<2.4$ and invariant masses in $10 <M_{\mu\mu}< 100$ GeV. The  STARLIGHT 1.1 calculations are in good agrement with the measured cross-sections across a wide range suggesting that the nuclear electromagnetic fields are reasonably described by the nuclear form factor and photon-fluxes used in the calculation. 

\section{ACKNOWLEDGMENTS}
Miguel Arratia was supported by Cambridge Trust, CONICYT Becas Chile 72140349 and Christ's College.

\nocite{*}
\bibliographystyle{aipnum-cp}%
\bibliography{sample}%

\end{document}